\documentclass[preprint,12pt]{elsarticle}

\usepackage{amssymb}

\usepackage{lineno}
\usepackage{upgreek}

\biboptions{numbers,sort&compress}

\journal{Nuclear Instruments \& Methods in Physics Research A}

\begin{document}

\begin{frontmatter}



	\title{Developing a mass-production model of large-area Si(Li) detectors
	with high operating temperatures}


	\author[JAXA]{M. Kozai\corref{corr}}
	\ead{kozai.masayoshi@jaxa.jp}
	\author[JAXA]{H. Fuke}
	\author[Shimadzu]{M. Yamada}
	\author[MIT]{K. Perez}
	\author[MIT]{T. Erjavec}
	\author[CU]{C. J. Hailey}
	\author[CU]{N. Madden}
	\author[MIT]{F. Rogers}
	\author[CU]{N. Saffold}
	\author[MIT]{D. Seyler}
	\author[KU]{Y. Shimizu}
	\author[Shimadzu]{K. Tokuda}
	\author[MIT]{M. Xiao}

	\address[JAXA]{Institute of Space and Astronautical Science,
	Japan Aerospace Exploration Agency (ISAS/JAXA), Sagamihara, Kanagawa 252-5210, Japan}
	\address[Shimadzu]{Sensor Device Business Unit, Device Department, Shimadzu Corporation,
	Atsugi, Kanagawa 243-0213, Japan}
	\address[MIT]{Massachusetts Institute of Technology, Cambridge, MA 02139, USA}
	\address[CU]{Columbia University, New York, NY 10027, USA}
	\address[KU]{Kanagawa University, Yokohama, Kanagawa 221-8686, Japan}

	\cortext[corr]{Corresponding author}

\begin{abstract}
	This study presents a fabrication process for lithium-drifted silicon (Si(Li))
	detectors that, compared to previous methods,
	allows for mass production at a higher yield,
	while providing a large sensitive area and low leakage currents
	at relatively high temperatures.
	This design, developed for the unique requirements of
	the General Antiparticle Spectrometer (GAPS) experiment,
	has an overall diameter of 10~cm, with ${\sim}$9~cm of active area
	segmented into 8 readout strips,
	and an overall thickness of 2.5~mm,
	with ${\gtrsim}2.2$~mm (${\gtrsim}90\%$) sensitive thickness.
	An energy resolution ${\lesssim}4$~keV full-width at half-maximum (FWHM)
	for 20$-$100~keV X-rays is required at the operating temperature
	${\sim}-40^\circ$C, which is far above the liquid nitrogen temperatures
	conventionally used to achieve fine energy resolution.
	High-yield production is also required
	for GAPS, which consists of ${\gtrsim}1000$ detectors.
	Our specially-developed Si crystal and custom methods of Li evaporation, diffusion and drifting
	allow for a thick, large-area and uniform sensitive layer.
	We find that retaining a thin undrifted layer on the $p$-side of the detector
	drastically reduces the leakage current,
	which is a dominant component of the energy resolution at these temperatures.
	A guard-ring structure and optimal etching of the detector surface are also confirmed
	to suppress the leakage current.
	We report on the mass production of these detectors that is ongoing now,
	and demonstrate it is capable of delivering a high yield of ${\sim}90\%$.
\end{abstract}


\end{frontmatter}


\section{Introduction}
We present here a high-yield mass production process
for lithium-drifted silicon (Si(Li)) detectors
that meet the unique requirements of the General Antiparticle Spectrometer (GAPS) experiment.
GAPS is a balloon-borne experiment that aims
to survey low-energy (${<}0.25$~GeV/n) cosmic-ray antinuclei for the first time,
by adopting a novel detection concept based on the physics of exotic atoms
\cite{Aramaki16a,Hailey09,Hailey13,Fuke17}.
Low-energy cosmic-ray antinuclei, especially antideuterons,
are predicted to be distinctive probes for
the dark matter annihilation or decay occurring in the Galactic halo
\cite{Donato00,Donato04,Duperray05,Aramaki16a,Cuoco17,Korsmeier18}.
Precise measurement of the low-energy antiproton spectra
will also provide crucial information on the source and propagation mechanisms
of cosmic rays \cite{Moskalenko02,Lin17,Bindi17,Aramaki14}.
GAPS sensitivities to antideuterons and antiprotons are discussed
in \cite{Aramaki16b} and \cite{Aramaki14},
and capabilities for antihelium detection are being evaluated.
The first flight of GAPS via a NASA Antarctic long duration balloon
is planned for late 2021.\par
GAPS is comprised of a $1.6\;\rm{m}^W \times 1.6\;\rm{m}^D \times 1.0\;\rm{m}^H$
tracker made of Si(Li) detectors surrounded by
a time-of-flight (TOF) system made of plastic scintillator paddles.
A low-energy antinucleus triggered by the TOF is
slowed and captured by the Si(Li) detector array,
forming an excited exotic atom with a silicon nucleus.
It immediately decays,
radiating de-excitation X-rays of characteristic energies.
The antinucleus then annihilates with the silicon nucleus,
producing pions and protons with a multiplicity
that scales with the incident antinucleus mass.
The surrounding Si(Li) detectors measure the energies
of the characteristic X-rays,
which are specific to the incident antinucleus species.
The ${\rm d}E/{\rm d}x$, trajectories, and total kinetic energies
of the incoming antinucleus and outgoing annihilation products,
as reconstructed by the Si(Li) and TOF systems,
provide additional antiparticle identification power.\par
Therefore, the Si(Li) detector array plays an essential role
in the GAPS detection scheme,
providing the absorption depth, active area, tracking efficiency, and
X-ray energy resolution necessary for this exotic atom particle
identification technique.
Our optimized flight detector design has an overall diameter of 10~cm,
with ${\sim}$9~cm of active area segmented into 8 readout strips,
and an overall thickness of 2.5~mm, with ${\gtrsim}2.2$~mm (${\gtrsim}90\%$)
sensitive thickness.
Both 4- and 8-strip designs have been developed
and proved to meet the GAPS requirement.
However, the 8-strip design has been adopted for the GAPS flight instrument,
as it improves tracking performance and allows use of a custom ASIC readout,
while still achieving the required energy resolution \cite{Rogers19}.
This low-power ASIC reduces inactive material in the tracker
and permits for lower temperature operation,
and thus improved noise performance,
compared to a discrete-component preamplifier readout.
The detectors are arranged in 10~layers each with ${\gtrsim}100$ detectors
to achieve the absorption depth necessary for incident antinuclei
with energies ${<}0.25$~GeV/n \cite{Aramaki16b,Aramaki14}.
Hence the mass production of ${\gtrsim}1000$ detectors is required.\par
The leakage current and capacitance of each Si(Li) strip should be
lower than 5~nA and ${\sim}$40~pF, respectively,
to achieve the required energy resolution
${\lesssim}4$~keV (FWHM) for 20$-$100~keV characteristic X-rays \cite{Rogers19}.
To suppress the power consumption,
GAPS adopts a newly-developed cooling system for the Si(Li) detectors \cite{Fuke16,Okazaki18}.
Considering all thermal and mechanical restrictions for a balloon-craft,
the cooling system is designed to cool Si(Li) detectors down to
${\sim}-40^\circ$C ($-35^\circ$C to $-45^\circ$C).
The leakage current, which depends exponentially on temperature,
must meet the requirement in this operating range.
The capacitance requirement can be simultaneously achieved by realizing a thick sensitive layer.
A relatively low bias voltage of 250~V is also required,
because it allows for operation at the ambient flight pressure
without suffering breakdown.\par
We have adopted Si(Li) detectors for the GAPS design, as they provide
a sensitive layer thicker than a few millimeters
with modest bias voltages,
by compensating $p$-type Si crystal with Li ions \cite{Pell60,Goulding66}.
However, a high-yield, low-cost fabrication method for large-area Si(Li) detectors,
with operating temperatures of at minimum ${\sim}-35$ to $-45^\circ$C,
currently has no experimental precedent.\par
Si(Li) detectors have been commercially-produced previously,
mainly in the energy dispersive X-ray (EDX) spectroscopy field.
However, all of these EDX Si(Li) detectors are small
with a diameter of ${\sim}1$~cm,
and operated mainly at liquid nitrogen temperature,
which is significantly lower than that required for GAPS.
Several previous studies have reported on large-area Si(Li) detectors
\cite{Miyachi88,Miyachi94,Kashiwagi90};
however, all these methods required removing the un-drifted region
after the Li drift,
because achieving a uniform drift across a large-area was impossible.
As an approach to address this difficulty,
it was demonstrated that uniform Li-drifting can be conducted into
a large-area Si wafer by drifting towards a boron-implanted $p$-side layer
\cite{Onabe02,Tindall04,Protic02,Protic03,Protic05}.
However, boron implantation requires annealing at ${\sim}500^\circ$C,
which has the potential to damage the Si crystals and increases
the fabrication cost.\par
\begin{figure}
	\begin{center}
	\includegraphics[width=0.5\textwidth,bb=200 0 700 400]{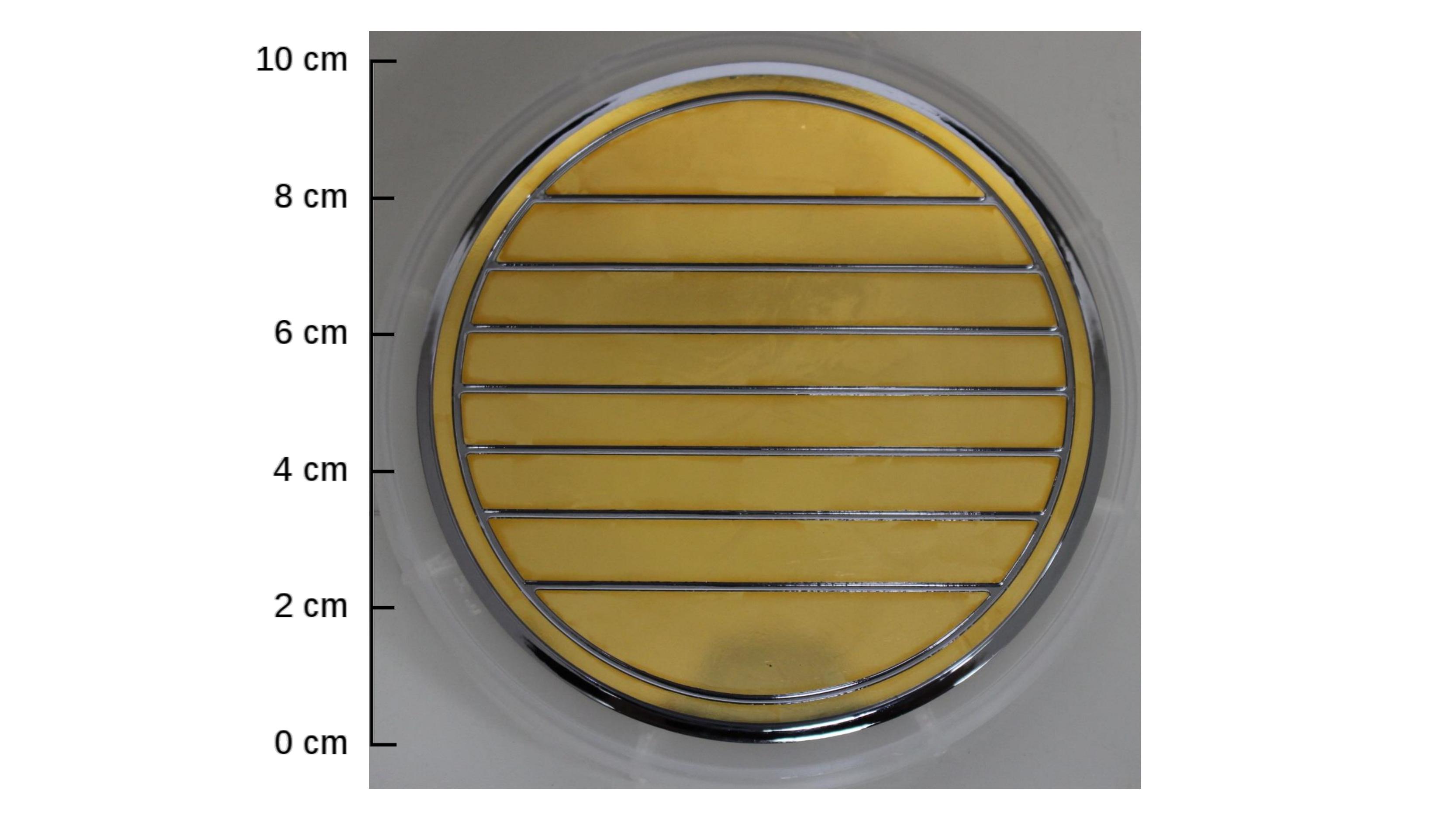}
		\caption{A GAPS Si(Li) detector with a thickness of 2.5~mm, diameter of 10~cm,
		and 8 readout strips.}
	\label{fig:det_pic}
	\end{center}
\end{figure}

Two key techniques in the GAPS Si(Li) development are
a uniform Li drift in large-area Si(Li) and
a suppression of leakage current at the relatively high temperature.
Based on previous research on prototype detectors
\cite{Aramaki10,Aramaki12,Perez13,Perez18},
we have established a mass-production method for the GAPS Si(Li) detectors.
Figure~\ref{fig:det_pic} shows an example of a GAPS Si(Li) detector.\par
In this paper, Sec. \ref{sec:overview} briefly explains the overall fabrication flow
of the GAPS Si(Li) detector.
Detailed descriptions of the fabrication process are given in Sec. \ref{sec:dev},
where we highlight the specific developments investigated
in over 50 prototype detectors.
Sec. \ref{sec:performance} demonstrates the performance and high-yield rate
of the mass-produced detectors.
Conclusions are presented in Sec. \ref{sec:conclusion}.

\section{Overview of fabrication flow}\label{sec:overview}
\begin{figure}
	\begin{center}
	\includegraphics[width=\textwidth,bb=30 100 550 550]{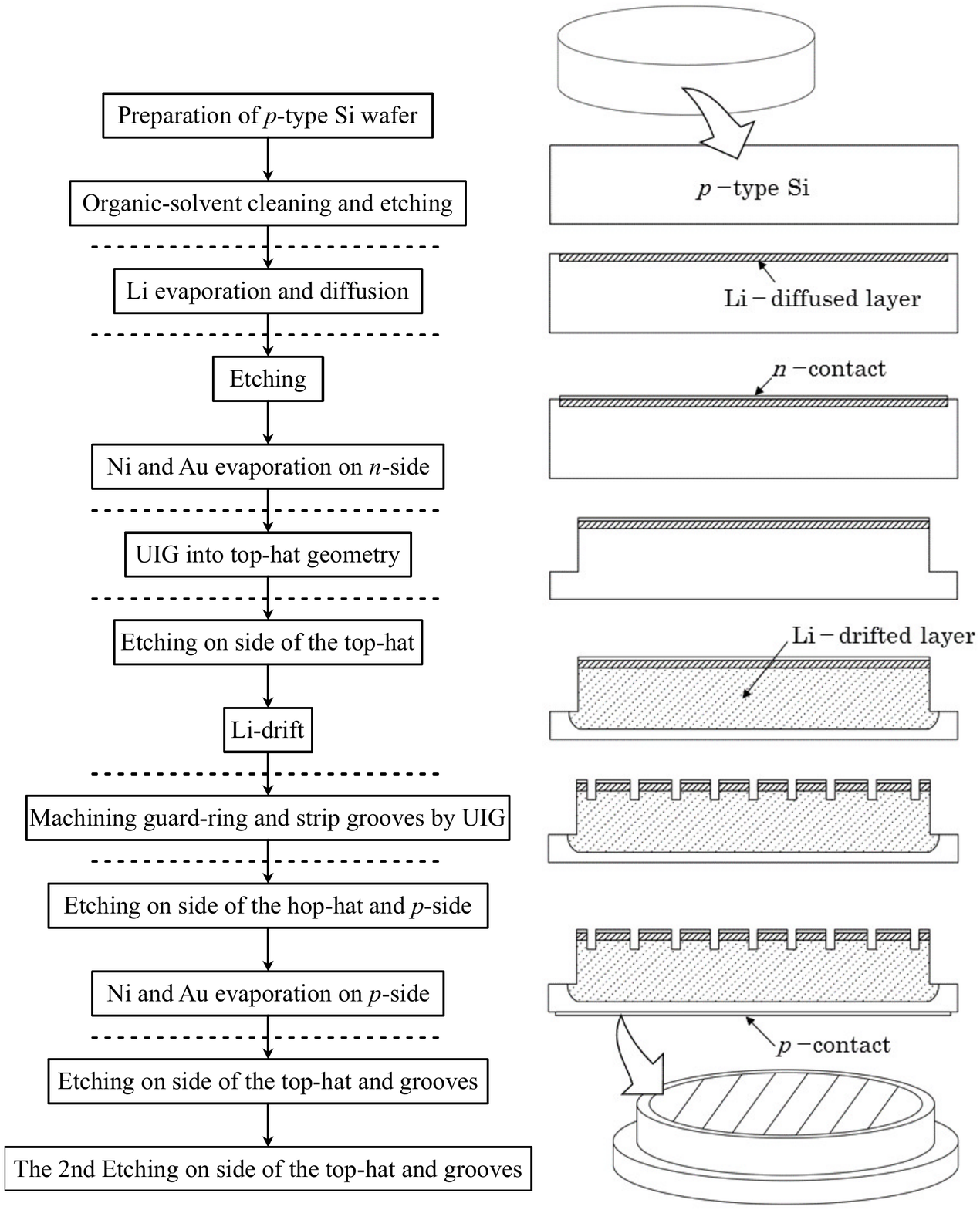}
		\caption{Fabrication flow of the GAPS Si(Li) detector.
	   The image on the right shows the cross-section of
		a Si(Li) detector in each process.
		Dimensions of the cross-section diagram are not to scale.}
	\label{fig:flow}
	\end{center}
\end{figure}
Figure~\ref{fig:flow} provides an overview of the fabrication process.
First, lithium is evaporated on one surface of a $p$-type Si wafer
and diffused through a shallow depth by heating the wafer.
This process supplies Li ions for compensating impurities in the $p$-type Si,
while forming a Li-diffused $n^+$-layer.
A metal contact is evaporated on $n$-side after the diffusion.
The wafer is machined into the top-hat geometry by
ultrasonic impact grinding (UIG) to confine the Li-drifted region.
Li ions in the Li-diffused layer are drifted toward the $p$-side
by applying a bias voltage to the heated wafer.
During drifting, Li ions compensate impurities in the $p$-type Si,
forming a well-compensated intrinsic layer ($i$-layer)
that functions as the sensitive volume in the final detector.
Grooves are machined on the $n$-side via UIG to isolate
the guard-ring structure and readout strips.
A metal contact is evaporated on the $p$-side.
The GAPS Si(Li) detector is then completed
by performing wet etchings and cleanings of its exposed Si surfaces.\par
\begin{figure}
	\begin{center}
	\includegraphics[width=\textwidth,bb=150 150 700 600]{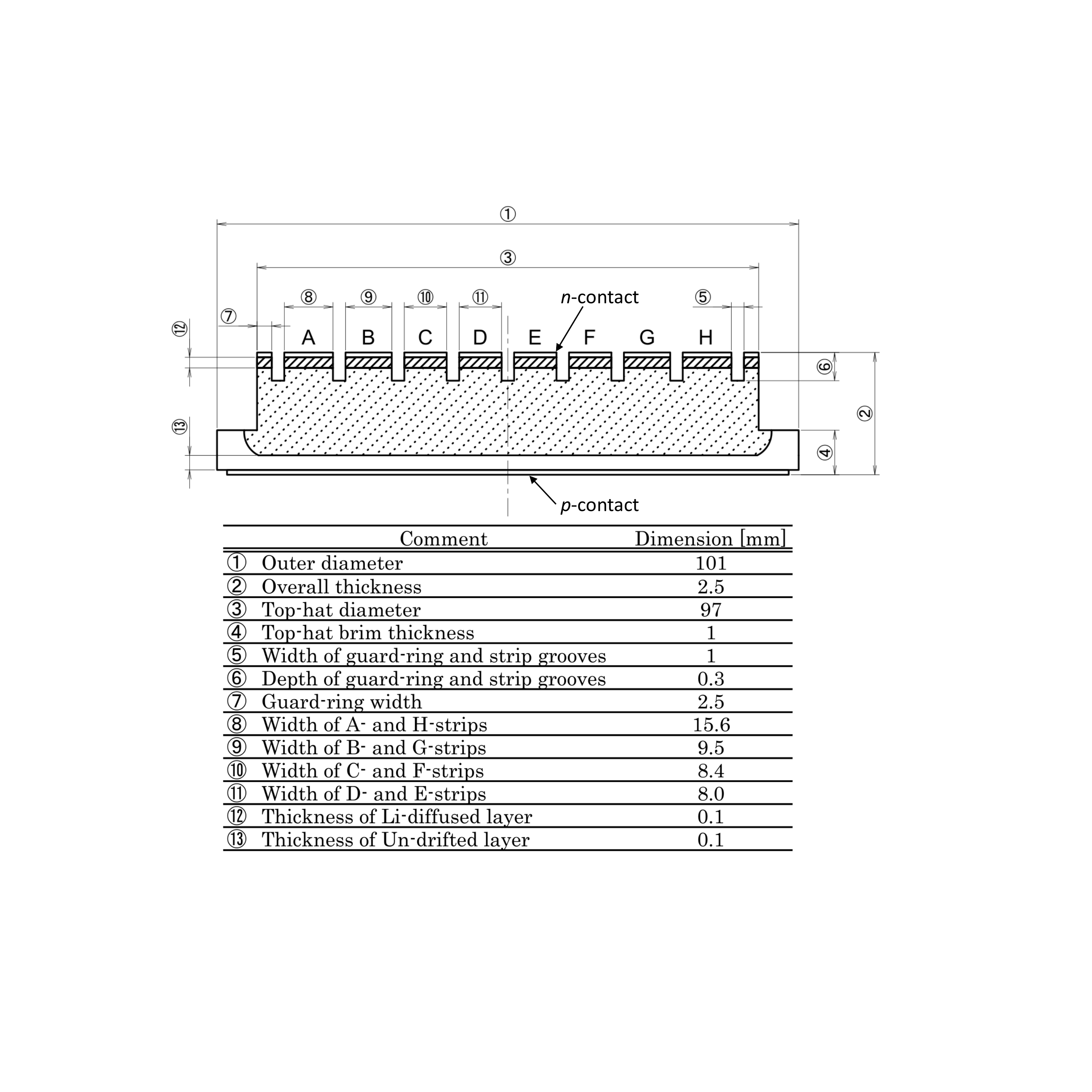}
		\caption{Half-section (upper figure) and its machined dimensions (lower table)
		of the GAPS Si(Li) detector.
		The cross-section figure is not to scale.}
	\label{fig:dim}
	\end{center}
\end{figure}
We summarize the dimensions of the mass-production model
of the GAPS Si(Li) detector in Fig.~\ref{fig:dim}.
Displayed dimensions represent the machined dimensions by UIG.
The final detectors deviate somewhat from these values
mainly due to material removed during etching.\par
Our developments prioritize cost effectiveness for mass production
while maximizing detector quality.
Shimadzu Corporation, Japan, is an industry-leading company of Si(Li) EDX detectors.
The fabrication apparatuses and methods herein are developed by
extending the work of our previous studies
\cite{Aramaki10,Aramaki12,Perez13,Perez18},
while utilizing Shimadzu's techniques
for quality control and cost reduction.

\section{Development}\label{sec:dev}
\subsection{Procurement of p-type Si crystal}
\begin{table}
	\begin{center}
	\caption{Specifications of Si crystal used for GAPS Si(Li) detector.}
	\begin{tabular}{ll}
		\hline
		Fabrication method & Floating zone \\
		Type & p \\
		Dopant & Boron \\
		Crystal orientation & $\langle111\rangle$ \\
		Oxygen concentration & $<1 \times 10^{16} \rm{\;atoms/cm^3}$ \\
		Carbon concentration & $<2 \times 10^{16} \rm{\;atoms/cm^3}$ \\
		Resistivity & ${\sim}1000 \rm{\;\Omega \cdot cm}$ \\
		Minority carrier lifetime & ${\sim}1$ ms \\
		Diameter & ${\sim}100$ mm (4 inches) \\
		\hline
	\end{tabular}
	\label{tbl:crystal}
	\end{center}
\end{table}
Several studies have reported
that uniformly drifting Li ions into a thick, large-area, $p$-type Si wafer is difficult
mainly due to defects and contaminants, such as oxygen and carbon, in the Si crystal
\cite{Miyachi88,Miyachi94,Kashiwagi90,Onabe02,Murray66,Fong82,Litovchenko03}.
These behave as traps for Li ions and
hence decrease Li ion mobility in the crystal, making it hard to uniformly drift.
Previous works also demonstrate that these impurities and defects
concentrate especially in the central region of Si wafer,
resulting in an uncompensated region
at the center of the wafer's $p$-side after the Li drift \cite{Kashiwagi90,Miyachi94,Onabe02}.\par
We have successfully developed a high-purity $p$-type Si crystal
specifically for the GAPS Si(Li) detectors in collaboration with SUMCO Corporation, Japan.
Table \ref{tbl:crystal} lists the specifications of the crystal used for our fabrications.
For the raw material of our crystal growth, we employ polycrystalline silicon made from mono-silane.
The crystal is grown to be oxygen free using the floating zone method
with an axis of $\langle111\rangle$.
Both Si crystals with $\langle111\rangle$ and $\langle100\rangle$ orientations are
used in previous studies,
but $\langle111\rangle$ is more proven for Si(Li) detector fabrication,
including for Shimadzu's commercial detectors.
There is also one report indicating that it is empirically
preferable for Li drift \cite{Kashiwagi90}.\par
Resistivity of ${\sim}1000 \rm{\;\Omega \cdot cm}$ corresponds to
an acceptor concentration of $N_A \approx 10^{13} \rm{\;atoms/cm^3}$ \cite{Goulding66},
which is an order of magnitude lower density than
that used in some of previous studies of large-area Si(Li) detectors
\cite{Miyachi88,Miyachi94,Kashiwagi90,Onabe02}.
Substrate with a lower $p$-type acceptor concentration requires fewer Li ions for compensation,
thus reducing the temperature and time required in the diffusion process.
Reduction of the heating treatment prevents the Si crystal from forming defects.
The in-plane non-uniformity of the resistivity of our Si wafer is ${\sim}10 \%$
based on the measurement of sample wafers.\par
The lifetime of minority carriers is an indicator of crystal defects and contaminants.
The lifetime of ${\sim}1$~ms is enough to make a high-quality compensated region,
as proven by our in-house development \cite{Perez18}.\par
After procuring 2.5~mm-thick wafers cut from the 10~cm-diameter Si crystal,
we remove foreign matter on the surface
by organic-solvent cleaning with methanol, xylene, and acetone.
We then etch the wafer surface with our etchant
(a solution of hydrofluoric acid, nitric acid and acetic acid)
for 2~min to remove surface contaminants and mechanical defects.
Si oxide on the surface is then removed by immersing the wafer
in a solution of hydrofluoric acid for 1~min.

\subsection{Li evaporation and diffusion}
Li is evaporated and diffused through a shallow depth, forming the $n^+$-layer.
The Si wafer is set in a chamber of a custom vacuum-based thermal Li evaporator
by sandwiching it between a heater plate and a lower mask.
The mask has a large round aperture that
prevents Li from being deposited onto the other side of the wafer.
Our custom heater plate realizes in-plane uniformity for heating the large-area wafer,
which is a key issue for obtaining a uniform Li-diffused layer.
A molybdenum evaporation boat filled with Li pieces is set under the wafer.
The shutter between the wafer and the boat is closed at the beginning of the process
to prevent Li oxide, on the surface of the Li pieces, from evaporating onto the wafer.
The chamber is then pumped to ${<}10^{-4}$~Pa, and the heater is set to 280$^\circ$C.
After confirming pressure and thermal stability,
the shutter is opened and the current through the boat is increased.
Li is evaporated onto the wafer surface by maintaining the current for 1~min,
and the pressure and temperature are kept for another minute
to diffuse Li through the shallow depth.
The pressure valve is then opened.
The wafer is naturally cooled to ${\lesssim}50^\circ$C, and extracted from the chamber.
We obtain a $n^+$-layer with a thickness of ${\sim}100\;\upmu\rm{m}$ by this diffusion process,
as confirmed by our visual inspection of the cross-section of the Si(Li) detector
via copper staining (see Sec. \ref{sec:drift}).\par
The depth of the $n^+$-$p$ junction formed by Li diffusion for time $t$ is expressed as \cite{Goulding66}
\begin{equation}
	x_j = 2\sqrt{D t} \cdot \rm{erfc}^{-1} \left( \frac{N_A}{N_0} \right) \label{eq:xj},
\end{equation}
where $D$ is the diffusion constant of Li in the Si crystal and $N_0$ is the Li surface concentration
($N_0 \approx 10^{16} \rm{\;atoms/cm^3}$ based on measurement of commercial Si(Li) \cite{Shimadzu16}).
The diffusion constant is obtained by \cite{Pell60} with
${\sim}1000 \rm{\;\Omega \cdot cm}$ $p$-type Si crystal as
\begin{equation}
	D = 6 \times 10^{-4} \exp \left(\frac{-0.61q}{k_{\rm B} T} \right) \rm{\;[cm^2/s]}, \label{eq:D}
\end{equation}
where $q$ is the elementary charge, $k_{\rm B}$ is the Boltzmann constant, and $T$ is temperature in K.
Our diffusion time $t$
(1${\sim}$2~min at 280$^\circ$C and
${\sim}$150~min of the natural cooling time)
leads to $x_j$ which is consistent with the practical depth ${\sim}100\;\upmu\rm{m}$,
though the empirical equation (\ref{eq:D}) of the diffusion constant is
derived or referred with different coefficients in each previous work
\cite{Pell60,Kashiwagi90,Fuller54,Maita58} and provides only an approximation.\par
Reference \cite{Pell60} indicates that the diffusion constant should decrease with lower resistivity,
i.e., highly-doped silicon crystal, by the ion-pairing effect.
This implies our high-purity Si crystal allows us to obtain a Li-diffused layer thick enough
with a lower temperature and a shorter heating time,
preventing defects from being generated in the crystal by our heating procedure.
Indeed, our heating temperature and time during the diffusion are
cooler and shorter than previous research
on large-area Si(Li) using ${\sim}100 \rm{\;\Omega \cdot cm}$ crystal.
For example, \cite{Miyachi88} and \cite{Miyachi94} maintained the wafer at 400$^\circ$C for 15~min
to form a ${>}100 \;\upmu \rm{m}$ diffused layer.
Reference \cite{Kashiwagi90} requires the wafer to be maintained at a temperature ranging
300$^\circ$C to 400$^\circ$C for 5$-$20~min
to obtain a diffused depth of 100$-$200~$\;\upmu \rm{m}$.\par
The Li oxide layer formed on the wafer surface is removed by chemical etching for 1~min
with our etchant.
This etching also removes mechanical defects and contaminants from the surface.

\subsection{Evaporation of {\it n}-electrode and top-hat machining}\label{sec:ncontact}
In a thermal evaporator set at room temperature and pressure of ${<}10^{-4}$~Pa,
an 18~nm layer of nickel followed by a 120~nm layer of gold are evaporated
onto the $n$-side of the detector to form a metal contact.
Nickel has good adhesion with the underlying Si, and gold has a high tolerance to oxidation.\par
The circumference of the {\it n}-side is then ground by UIG
to make a top-hat geometry (see Fig.~\ref{fig:flow}) and
to confine the Li from drifting to the sides of the wafer
during the Li drifting procedure.
UIG is generally less expensive than other methods such as
diamond-saw cutting,
and is thus more suitable for mass production.\par
The geometry of the top-hat brim is designed to be as narrow as possible
to maximize the detector's active area,
while still being wide enough to be used for handling the detector.
The top-hat brim is preferred for handling the detector
since it is $p$-type even after the Li drift and thus does not need careful treatment
(unlike the intrinsic region ($i$-region) surface, see Sec. \ref{sec:groove}).
As a result of this optimization,
the inner diameter of the top-hat brim is set to 97~mm in the UIG machining,
and the thickness is set to 1~mm, to prevent breakage during handling.\par
After the top-hat machining,
an etch-resisting wax, Apiezon$^{\scriptsize\textregistered}$, is dissolved in xylene
and is painted on the $n$- and $p$-side surfaces by hand.
Damaged layers and metal contaminants generated by UIG are
then removed by etching on the side surface of the top-hat for 12~min.
Organic-solvent cleaning with methanol, xylene, and acetone is then performed to remove the wax.

\subsection{Li drift}\label{sec:drift}
\begin{figure}
	\begin{center}
	\includegraphics[width=1.3\textwidth,bb=0 0 300 200]{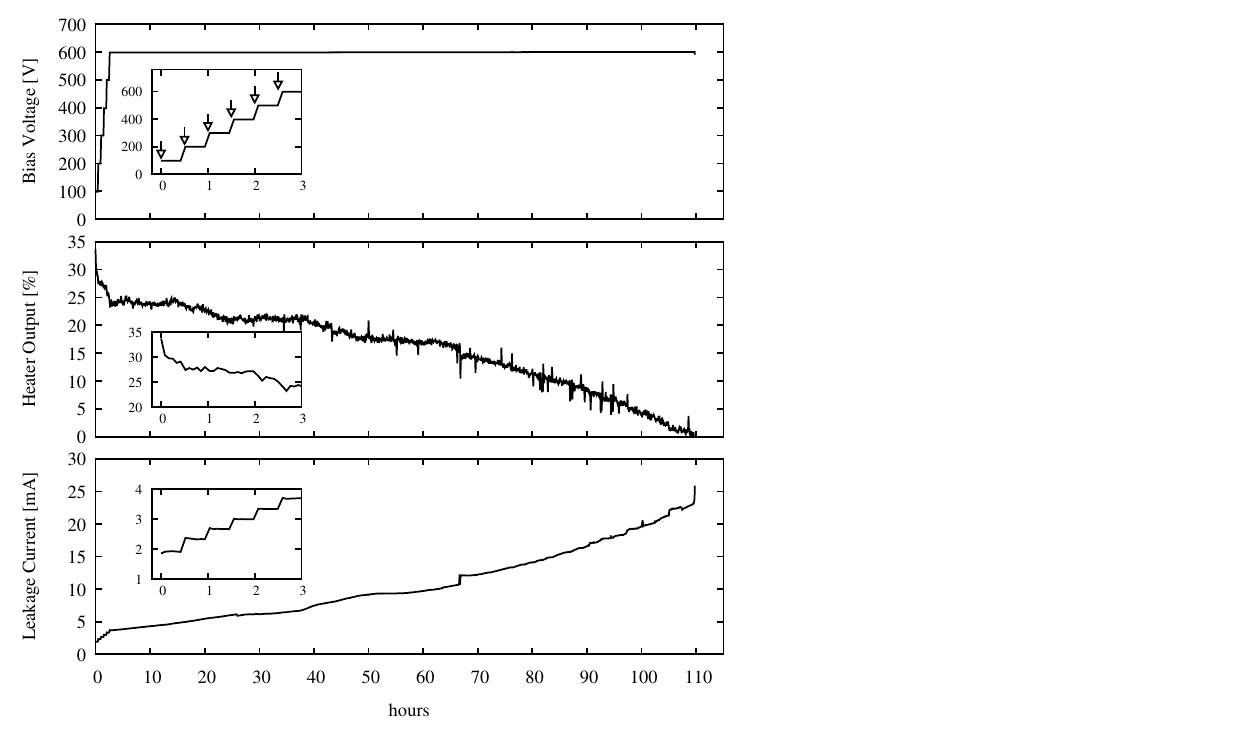}
		\caption{Profiles of the applied bias voltage (top), heater output (middle),
		and leakage current (bottom) during the drift of a sample detector.
		The inset plot in each panel displays each parameter's variation in the first 3~hours of the drift,
		as the voltage is increased step-wise to the 600~V set-point.}
	\label{fig:drift}
	\end{center}
\end{figure}
A uniform Li-drift is achieved by our drift apparatus,
which has been custom-designed for large-area Si(Li) detectors.
We found that retaining a thin undrifted layer on the $p$-side
effectively suppresses the leakage current, as demonstrated in this section.
The radial uniformity of the growth of the drifted region
during Li drift is key to realizing this thin, uniform undrifted layer.\par
Li drifting is performed in a custom drift apparatus consisting of
an electrically grounded heater plate,
a pressure contact for applying a bias voltage,
a resistance temperature detector (RTD),
and a computer-based controller.
The Si wafer is set on the heater plate with the $n$-side up and,
the $p$-side connected to the grounded heater plate.
The pressure contact and RTD are connected to the $n$-electrode
to apply a positive bias voltage and monitor the wafer's temperature.
The drift sequence described below is automatically managed
via the computer-based controller.\par
Figure~\ref{fig:drift} depicts an example of
the bias voltage, heater output, and leakage current during the ${\sim}$110~hours
of the drifting routine.
Our sample detector is a 10-cm Si(Li) detector.
The voltage is increased step-wise, in 100~V intervals every 30~min,
to prevent rapid increase of the leakage current.
The insets in each panel of Fig.~\ref{fig:drift} display the variations
over the first 3~hours of the drift,
as the voltage is ramped up to the set-point of 600~V.
In the first panel, open arrows indicate the timing of the voltage steps.
In the final panel, the leakage current shows a step-like increase of ${\sim}0.5$~mA
corresponding to each voltage increase.\par
As Li drifts toward the $p$-side and the depletion layer expands from the $n$-side,
the leakage current gradually increases.
The Joule heat generated by the leakage current also increases.
Displayed by the middle panel, heater output is automatically decreased
to compensate for the Joule heat and keep the wafer at 100$^\circ$C.\par
At the end of the drift, the depletion layer approaches the $p$-side.
At this point, the leakage current rapidly increases,
the wafer temperature exceeds 100$^\circ$C due to the Joule heat,
and the heater output decreases to zero.
The bias voltage is automatically turned off, i.e., the Li drift is terminated,
either when the leakage current reaches 25~mA or when the heater output becomes zero.
The wafer is then allowed to naturally cool to room temperature.\par
\begin{figure}
	\begin{center}
	\includegraphics[width=\textwidth,bb=0 0 900 500]{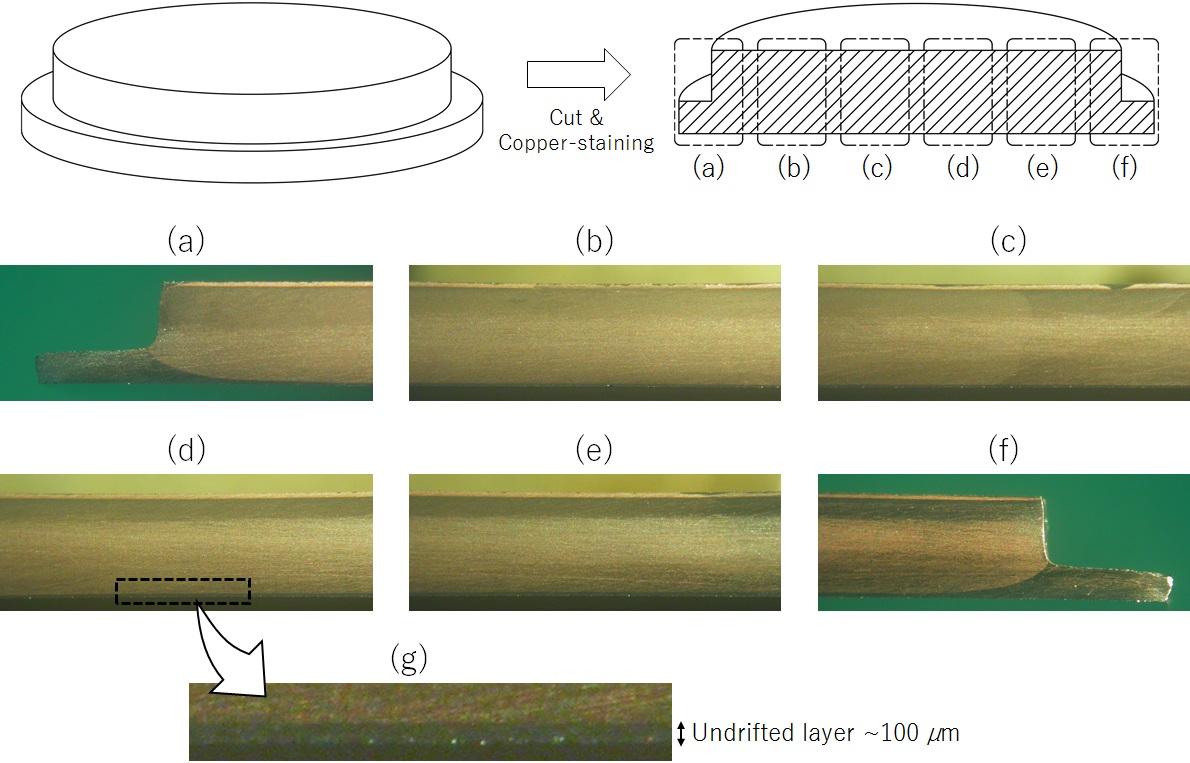}
		\caption{Six cross-sections of a sample Si(Li) detector
		after Li-drifting and copper staining.
		Sections (a)-(f) are indicated in the top figure.
		Section (g) displays an enlarged view of the indicated region in section (d)
		highlighting the undrifted layer.}
	\label{fig:copper}
	\end{center}
\end{figure}

Under a bias voltage as high as 600~V,
the depletion layer expands slightly toward the $p$-side beyond
the ``$i$-$p$ junction'' formed between the drifted and undrifted layer.
Thus, despite the steep increase in leakage current,
a thin undrifted layer is retained on the $p$-side of our Si(Li) detectors
after the drifting process, as demonstrated in Fig.~\ref{fig:copper}.\par
The necessary drift time ($t$) to obtain a drifted depth ($W$) under bias voltage ($V$) is given
by \cite{Goulding66}
\begin{equation}
	t = \frac{W^2}{2V\mu_L}. \label{eq:tdrift}
\end{equation}
The Li mobility ($\mu_L$) is related to the diffusion constant ($D$) by the Einstein relation
\begin{equation}
	\mu_L = \frac{q}{k_{\rm B} T} D \;[\rm{cm^2/(V \cdot s)}]. \label{eq:mu}
\end{equation}
For our drift parameters,
we derive $t \approx 100$~h to obtain a drifted depth $W \approx 2.2$~mm
(${\sim}90\%$ of the overall 2.5~mm thickness).
This calculated result is comparable to our actual drift time displayed in Fig.~\ref{fig:drift}.
In the case that the Li drift is unexpectedly terminated before ${\sim}90$~h,
we resume the drift sequence.\par
The temperature and voltage allow us to achieve a high quality drift
in an acceptable drift time (${\sim}4$~days) for mass production.
Higher bias voltage and temperature would reduce the required drift time.
However, setting the wafer at a higher temperature by higher Joule heat or higher heater output
will generate more hole-electron pairs, which attract Li-compensation,
hence disturbing the ideal Li distribution,
which should only compensate for acceptors \cite{Goulding66}.
Our drift parameters, 600~V and 100$^\circ$C, are found to be sufficiently low
to fabricate uniform and thick Li-drifted.\par
The Li-diffused and drifted regions can be visually inspected
by copper staining on a cross-section of the Li-drifted Si wafer \cite{Whoriskey58,Iles60,Kume01}.
In the copper-staining process, we first remove the $n$-electrode via etching and
cut the wafer into two cross-sectional pieces via diamond-saw dicing.
The cross-section is then polished and immersed in a saturated $\rm{CuSO_4}$ solution containing
a few drops of concentrated hydrofluoric acid under the illumination of white light.
Electrons generated by the light are more attracted to $n^+$- and $i$-regions than $p$-regions
due to the photovoltage effect and cause the deionization of copper ions,
resulting in copper deposition on the $n^+$- and $i$-regions.\par
The copper-stained area is visually identified by a yellow-ish color.
Figure~\ref{fig:copper} displays a cross-section of our sample detector processed
by this copper-staining method.
The light yellow-ish region corresponds to the Li-drifted region,
whereas the darker grey-ish region filling the brim of
the top-hat in Figs. \ref{fig:copper}(a) and \ref{fig:copper}(f) corresponds
to the undrifted $p$-type bulk.
The thin but intense yellow-ish layer on the $n$-side of the wafer indicates the $n^+$-layer,
i.e., the Li-diffused layer.
We confirm via this staining that the $n^+$-layer has a thickness of
${\sim}100 \;\upmu\rm{m}$.\par
The undrifted layer is also found in Fig.~\ref{fig:copper};
the dark grey-ish region seen at the brim spreads toward
the center and extends to other side of the brim,
keeping a thin undrifted region on the $p$-side.
Figure~\ref{fig:copper}(g) displays
an enlarged view of the $p$-side region in picture (d),
highlighting the undrifted layer.
This undrifted layer appears uniform, with a thickness of ${\sim}100 \;\upmu\rm{m}$
across what will become the active region of the detector,
indicating a successful uniform Li-drift.\par
\begin{figure}
	\begin{center}
	\includegraphics[width=0.75\textwidth,bb=0 0 360 190]{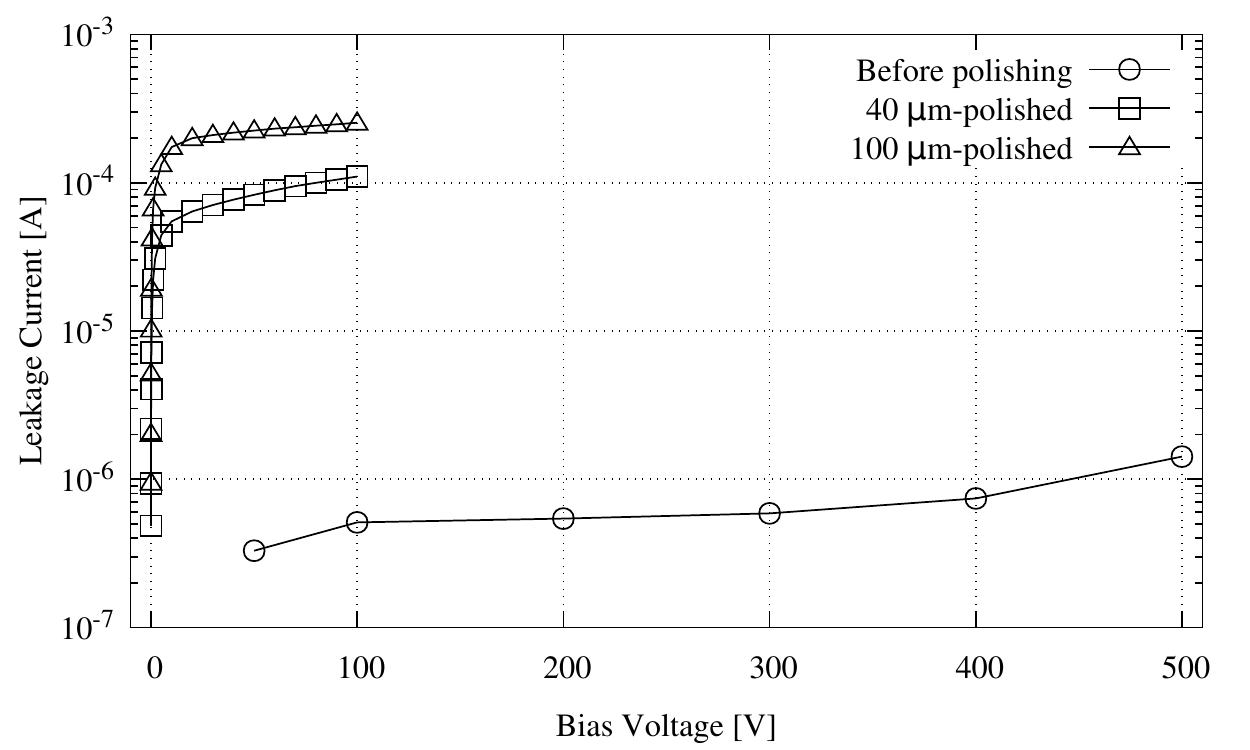}
		\caption{Leakage current of a sample detector as a function of bias voltage
		(I$-$V curve) at $-35^\circ$C before and after polishing undrifted layer on $p$-side.
		Circles display the I$-$V curve before polishing undrifted layer on $p$-side,
		whereas rectangles and triangles display the I$-$V curves after polishing $p$-side
		for depths of 40 and $100 \;\upmu\rm{m}$, respectively, from the unpolished $p$-side surface.}
		\label{fig:polish}
	\end{center}
\end{figure}
In conventional Si(Li) detector fabrication techniques,
the undrifted layer is removed to expose the Li-drifted layer on the $p$-side.
Then a metal contact such as gold is evaporated as a Schottky barrier.
However, we observed excessively high leakage currents at $-35^\circ$C
with a detector using this conventional contact.\par
Figure~\ref{fig:polish} displays the leakage current of our 10-cm diameter Si(Li) detector
at $-35^\circ$C as a function of bias voltage (I$-$V curve).
In this prototype detector, the guard-ring structure described in Sec. \ref{sec:groove},
which is necessary to suppress the dominant surface leakage current
and without which the effect of the undrifted layer is difficult to observe,
had already been machined. 
On the contrary, machining of the strip grooves and the second etching,
described in the Sections \ref{sec:groove} and \ref{sec:etching} respectively,
were not performed in this detector.
The circles in Fig.~\ref{fig:polish} display the I$-$V curve of the detector without polishing the $p$-side,
i.e., with the ${\sim}100 \;\upmu\rm{m}$ undrifted layer
remaining underneath the $p$-side nickel and gold contact.
This $p$-side contact is formed after etching the $p$-side surface,
in the same manner as Sec.~\ref{sec:pcontact}.
After measuring this I$-$V curve, the $p$-side was polished to remove 40~$\upmu$m,
leaving ${\sim}60 \;\upmu\rm{m}$ of undrifted layer,
and the nickel and gold $p$-side contacts were re-applied.
In this manner the thickness of the undrifted layer can be
reduced easily and precisely
in comparison to other options such as
modifying the undrifted thickness
by modifying the drifting parameters.\par
The rectangles in Fig.~\ref{fig:polish} show the I$-$V curve of this detector
after removing 40~$\upmu$m of the undrifted layer.
The leakage current is significantly higher than that of the unpolished detector
by an order of ${\sim}10^2$.
We next removed an additional 60~$\upmu$m from the $p$-side of the detector
via polishing,
thus removing the entire undrifted $p$-side layer,
and re-applied the nickel and gold contacts.
The resulting I$-$V curve is indicated by the triangles in Fig.~\ref{fig:polish}.
The metal contact should form a Schottky barrier on the $p$-side
but the leakage current is higher than that of the unpolished detector (circles)
by an order of ${\sim}10^3$.\par
This result implies that a large-area Schottky barrier contact easily breaks down
at our operating temperature (${\sim}-35$ to $-45^\circ$C),
which is significantly higher than
that of conventional Si(Li) detectors used for X-ray spectrometry.
Based on this presumption, we decided to retain the undrifted layer of
${\sim}100 \;\upmu\rm{m}$ on the $p$-side to suppress the bulk leakage current,
which we posit is generated in or flowing through the depletion layer via
a junction between the $i$- and $p$-layers.\par
In the conventional detectors,
the $p$-contacts are used as windows for low-energy X-rays \cite{Lyman89,Rossington91,Cox05}.
Therefore, such detectors adopted Schottky barrier contacts to minimize the $p$-side insensitive layers
while suppressing bulk leakage currents.
Since GAPS aims to detect X-rays with energies higher than 20~keV,
the ${\sim}100 \;\upmu\rm{m}$ insensitive layer is acceptable for the GAPS detectors.

\subsection{Machining grooves for the guard-ring and strips}\label{sec:groove}
\begin{figure}
	\begin{center}
	\includegraphics[width=0.75\textwidth,bb=0 0 360 190]{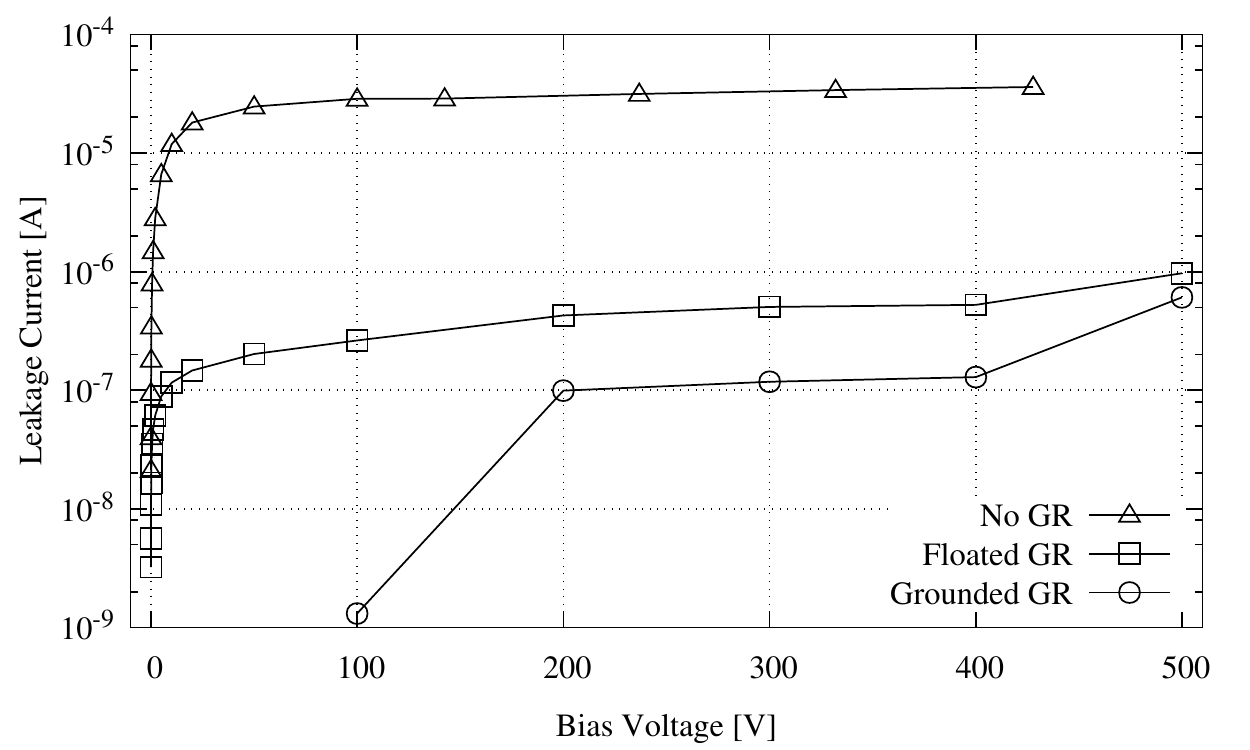}
		\caption{I$-$V characteristics of a sample detector at $-35^\circ$C
		before machining the guard-ring groove (triangles);
		after machining the groove, with a floated guard-ring (rectangles);
		and after grounding the guard-ring (circles).}
	\label{fig:GR}
	\end{center}
\end{figure}
The exposed $i$-region can be easily contaminated
and contribute to the leakage current \cite{Goulding66,Goulding61}.
The side surface of the top-hat has the largest area of the exposed $i$-layer and
is a major source of the leakage current.
The guard-ring groove \cite{Goulding61} is machined to suppress this surface leakage current,
preventing it from flowing into the readout electronics.\par
After the Li-drifting process, a circular groove, with a ${\sim}300 \;\upmu\rm{m}$ depth and
1-mm width, is cut into the $n$-side by UIG
so that the central area, or the readout electrode of the detector,
is electrically isolated from the perimeter, or guard-ring electrode (see Fig.~\ref{fig:det_pic}).
The groove is deep enough to cut through the $n^+$-layer (${\sim}100 \;\upmu\rm{m}$ depth).
During detector operation with a high bias voltage,
the $i$-layer between the guard ring and the active area is fully depleted and
its resistivity increases dramatically
to isolate the readout electrode from the guard-ring electrode.\par
Figure~\ref{fig:GR} displays the I$-$V curves of a sample detector.
The undrifted layer described in Sec. \ref{sec:drift} had been formed in this prototype detector,
but the machining of the strip grooves (described below) and
the second etching (described in Sec. \ref{sec:etching}) were not performed.
The I$-$V curves in Fig.~\ref{fig:GR} are measured
before machining the guard-ring groove (triangles);
after machining the groove, with a guard-ring not connecting to any external electrode (rectangles);
and after grounding the guard-ring (circles).
As described in Sec. \ref{sec:performance},
negative bias voltage is applied to the $p$-electrode, and
the strips that are not being measured are connected to the ground.
As shown in Fig.~\ref{fig:GR}, the grounded guard-ring
drastically reduces the leakage current flowing into the readout electrode
by an order of ${<}10^{-2}$ in comparison to the detector without a guard-ring.
This indicates that the readout electrode is successfully isolated
from both the guard-ring electrode and the side surface of the detector.\par
Grooves that divide the active area into 8 readout strips of equal area are cut
at the same time as machining the guard-ring groove,
using the same groove depth and width.
Under the operating bias voltage,
the strips are electrically isolated from each other
by the same principle as the guard-ring.\par
A narrower groove width is preferable
because the exposed $i$-layer in the groove can generate leakage current.
However, it is important to note that UIG is limited by the width of its grinding tool.
An extremely thin tool is difficult to make and is easily damaged in the UIG process.
A narrower groove also has disadvantages for groove etchings.
If the groove is too narrow,
bubbles generated via chemical reactions in the etching process stick to the groove surface.
The bubbles make it difficult to etch uniformly and sufficiently.
We optimized the UIG tool for the mass-production model of the GAPS Si(Li) detector and
deduced the optimal groove width as 1~mm.\par
The guard-ring width, i.e., width of the perimeter electrode in Fig.~\ref{fig:det_pic},
is also optimized.
A narrow guard-ring electrode is preferable to make the active area as wide as possible.
However, it is difficult to precisely paint etch-resisting wax,
which is necessary during the groove etching process, onto a narrow guard ring.
We deduced that the optimal guard-ring width is 2.5~mm.

\subsection{Forming $p$-side contact}\label{sec:pcontact}
Wax is painted on the $n$-side including the grooves.
Then, the side surface of the top-hat and the $p$-side surface are etched for 1~min.
After the wax is removed by an organic solvent,
the metal contact on the $p$-side is evaporated in the same manner as Sec. \ref{sec:ncontact}.

\subsection{Etching on side of the top-hat and grooves}\label{sec:etching}
\begin{figure}
	\begin{center}
	\includegraphics[width=0.75\textwidth,bb=0 0 360 200]{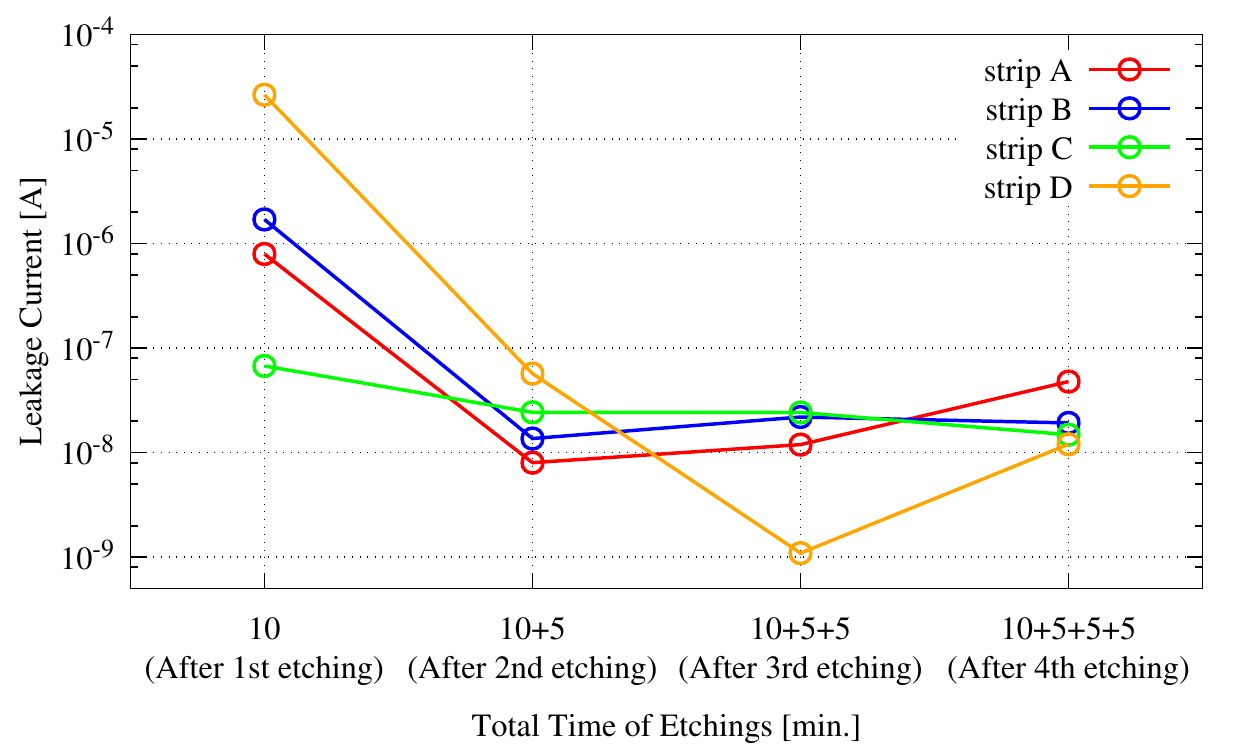}
	\caption{Leakage current of each strip
		at a bias voltage of 200~V and a temperature of $-35^\circ$C
		after the 1st, 2nd, 3rd, and 4th $n$-side groove etchings.
		The 1st etching is performed for 10~min,
		whereas the 2nd$-$4th etchings are performed for 5~min.
		The numbers below the horizontal axis show the cumulative etching time.}
	\label{fig:etching}
	\end{center}
\end{figure}
The side of the top-hat and the $n$-side grooves are etched
after painting wax on the $n$- and $p$-electrodes.
This last etching not only removes the damaged layer in the grooves
formed by UIG in Sec. \ref{sec:groove}
but also smooths the surfaces and removes contaminants
from all areas of exposed silicon.
Organic-solvent cleaning is performed after this etching to remove the wax.
It is assumed that this cleaning with methanol produces
a lightly $n$-type surface on the exposed $i$-region,
which ensures no electric breakdown of the detector
under the high bias voltage \cite{Goulding66,Llacer64}.\par
An optimized etching time is required
because etching for too long not only decreases the active area
but also expands the area in which the $i$-layer is exposed.
This area is easily contaminated, which increases leakage current.
The number of separate etchings is also a key parameter.
The wax is painted by hand,
creating a non-uniform line at the edge of the painted area
and thus introducing non-uniformities to the edge of the electrode after etching.
Bubbles generated in the etchant also make the surface state non-uniform.
Repeating the wax painting and the etching reduces these irregularities.\par
We surveyed the optimal etching time and
number of discrete wax painting and etching processes.
Figure~\ref{fig:etching} displays the leakage current of each strip of a prototype detector
with 4 strips (A$-$D)
at a bias voltage of 200~V and a temperature of $-35^\circ$C.
The leakage current is displayed as a function of
the cumulative etching time or the total number of etchings.
The first etching is performed for 10~min,
whereas the second, third, and fourth etchings are 5~min each.
Leakage currents in all strips are clearly reduced after
the second etching, i.e., cumulative etching time of 15~min,
compared to that after the first etching.
It was also confirmed by a visual inspection that the groove surfaces are
clearly smoothed and glassy after the second etching compared to after the first etching.
The leakage currents are not significantly affected after the additional third and fourth etchings.
This result indicates that the 15~min of etching in two discrete etching procedures
is sufficient to minimize the surface leakage current.

\section{Performance of the mass-produced detectors}\label{sec:performance}
\begin{figure}
	\begin{center}
		\includegraphics[width=\textwidth,bb=0 0 900 400]{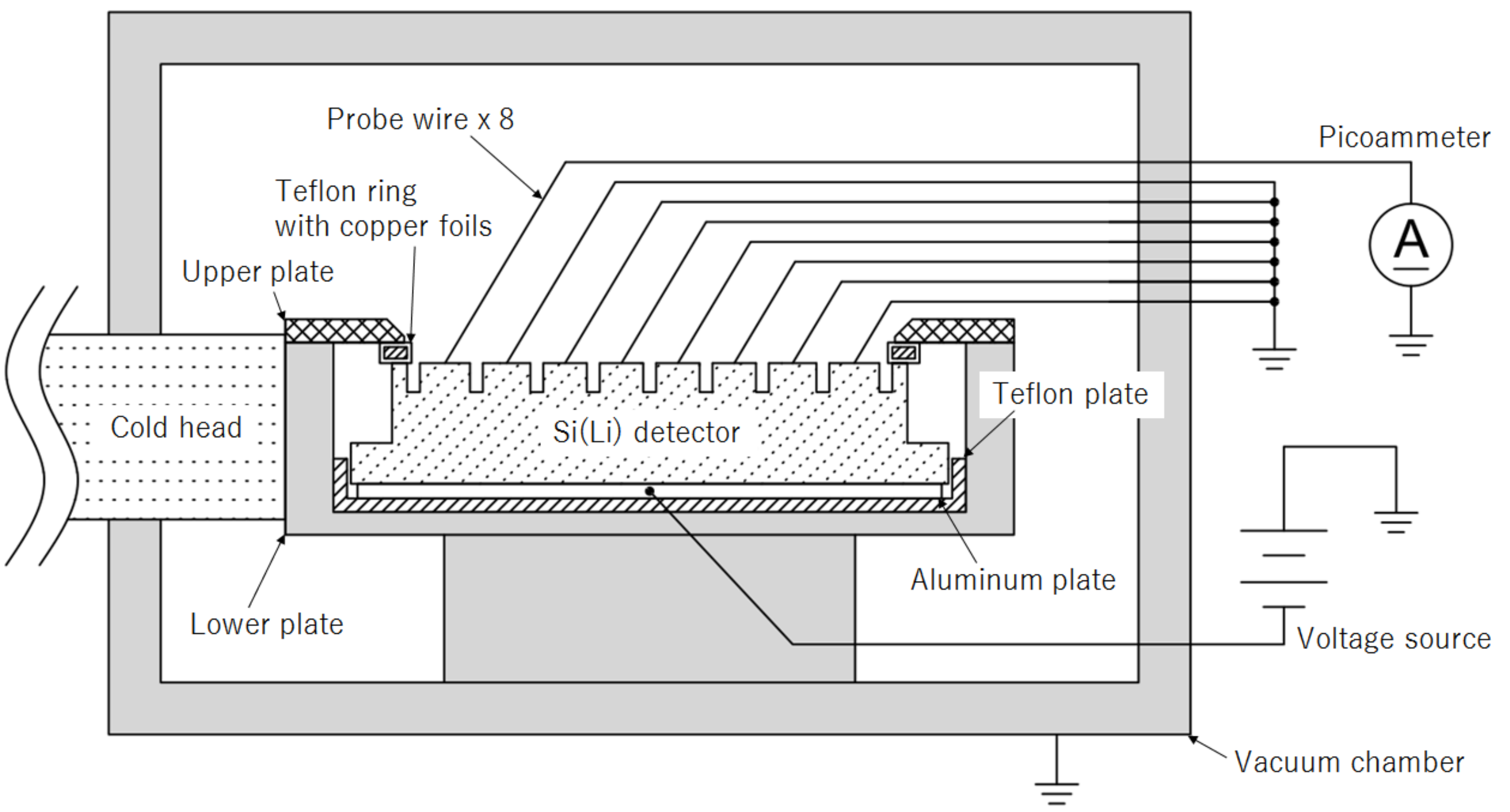}
		\caption{Diagram of the I$-$V curve measurement.}
		\label{fig:measure}
	\end{center}
\end{figure}
Ten prototype detectors (test detectors) were fabricated to validate
and fix the mass-production process.
Then, the mass production of the flight detectors had been started.
In this section, we evaluate leakage currents and capacitances
of these mass-produced detectors (10 test detectors and 10 flight detectors),
which were fabricated by identical procedures.
Thanks to the sufficiently low leakage currents,
it has been confirmed that energy resolution ${\lesssim}4$~keV (FWHM) for 20$-$100~keV X-rays
are achieved \cite{Kozai18}.
Details of the energy resolution measurements are described in a separate publication \cite{Rogers19}.\par
Figure~\ref{fig:measure} illustrates the setup for the I$-$V curve measurement.
The detector is mounted in a vacuum chamber set at $10^{-2}$~Pa.
A Teflon plate mounted on the lower plate of the detector holder
electrically isolates the $p$-electrode of the detector from the lower plate,
which is fixed to the chamber and connected to the ground.
An aluminum plate between the Teflon plate and the detector
applies negative bias voltage to the $p$-electrode.
The upper plate of the holder is bolted to the lower plate,
holding in place a Teflon ring pressed against the guard-ring.
The detector is fixed by this pressure.
Copper foils are pasted on the Teflon ring
to electrically connect the guard-ring and the upper plate;
therefore, the guard-ring electrode is grounded via the chamber.\par
A cold head bolted to the detector holder is cooled by a Stirling cooler.
The detector is cooled by the thermal connections to the holder
through the copper foils, Teflon ring,
alminum plate, and Teflon plate.
Calibration of the cooler was performed using
a dummy wafer on which an RTD was used to monitor the detector's temperature.\par
All strips of the detector are connected
to the hermetic connector mounted on the side of the chamber
via probes made of gold wires.
The gold wire probe is selected to prevent damage to the strip electrodes on the detector.
Using a multimeter, we verify the electrical connections between the hermetic connector and strips
and reasonable resistivities between the strips, guard-ring, and $p$-electrode,
each time the detector is set.
The strip to be measured is connected to a picoammeter or capacitance meter while
the other strips are connected to the ground outside the chamber.\par
For the picoammeter, capacitance meter, and bias supply,
HP 4140B, HP 4280A, and ORTEC 428 were used, respectively.
Recently the picoammeter and bias supply were replaced with a Keithley 6487 picoammeter,
which has an internal bias supply.
We constructed an automatic I$-$V measurement system
controlled by LabVIEW software with this Keithley picoammeter.
It will improve the speed of the mass production of the GAPS flight detectors
in terms of detector evaluation.\par
\begin{figure}
	\begin{center}
	\includegraphics[width=0.8\textwidth,bb=100 0 345 180]{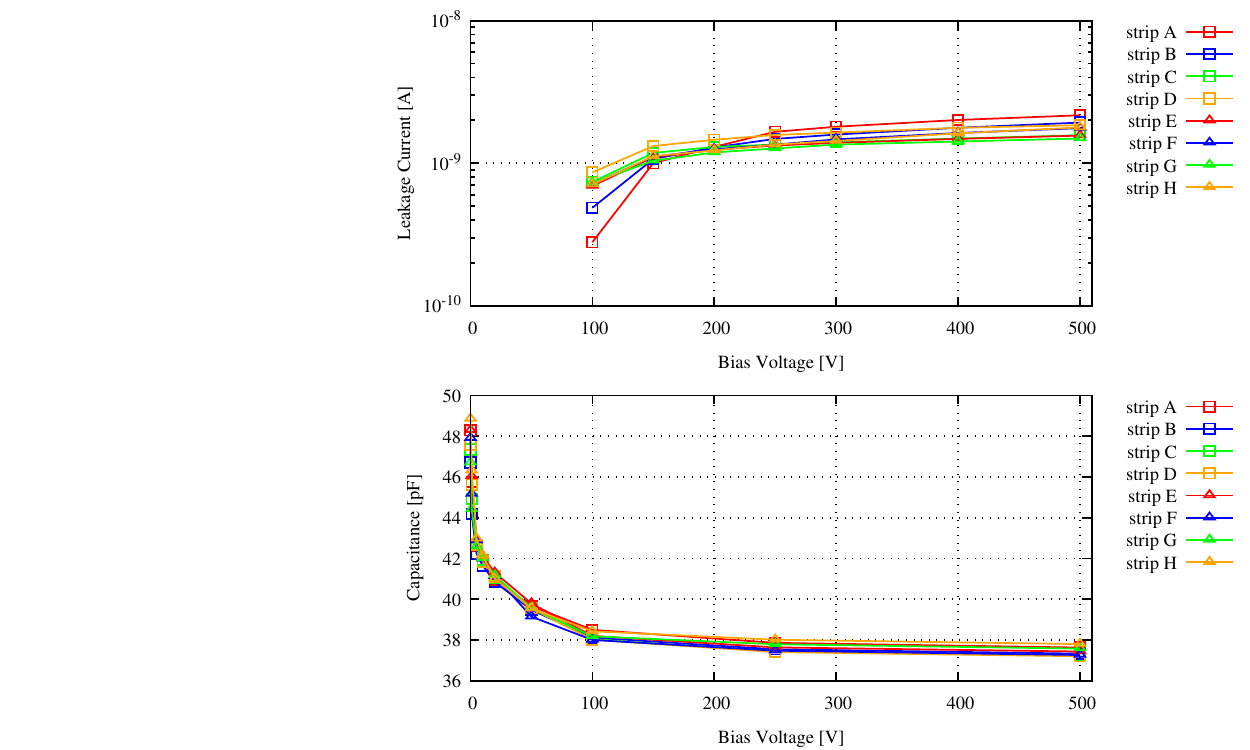}
		\caption{Leakage currents (upper panel) and capacitances (lower panel)
		of all strips (A$-$H) of a sample detector at $-35^\circ$C,
		as functions of the bias voltage.}
	\label{fig:IC}
	\end{center}
\end{figure}
The upper panel of Fig.~\ref{fig:IC} displays I$-$V curves of 8 strips
of a sample detector at $-35^\circ$C.
The leakage current of each strip is stable and is ${\sim}1.5$~nA
at our operating voltage (200$-$300~V),
significantly lower than the requirement, ${\sim}5$~nA \cite{Rogers19}.
The capacitance as a function of the bias voltage in the lower panel of Fig.~\ref{fig:IC}
reaches the minimum by ${\sim}100$~V,
indicating the detector is fully depleted around this voltage.
The capacitance of $C \approx 38$~pF also meets the requirement and
corresponds to the thickness of the depletion layer \cite{Goulding66},
\begin{equation}
	W_d \approx 1.05 A/C \approx 2.2 \;\rm{mm},
\end{equation}
for a strip area of $A=8.1 \;\rm{cm^2}$.
This is ${\sim}90\%$ of the 2.5~mm total detector thickness.
The uniformity of the capacitance, with a fluctuation of ${\lesssim}{\pm}1.5\%$
between strips, indicates a uniform Li-drifted layer.\par
\begin{figure}
	\begin{center}
	\includegraphics[width=0.8\textwidth,bb=0 0 320 200]{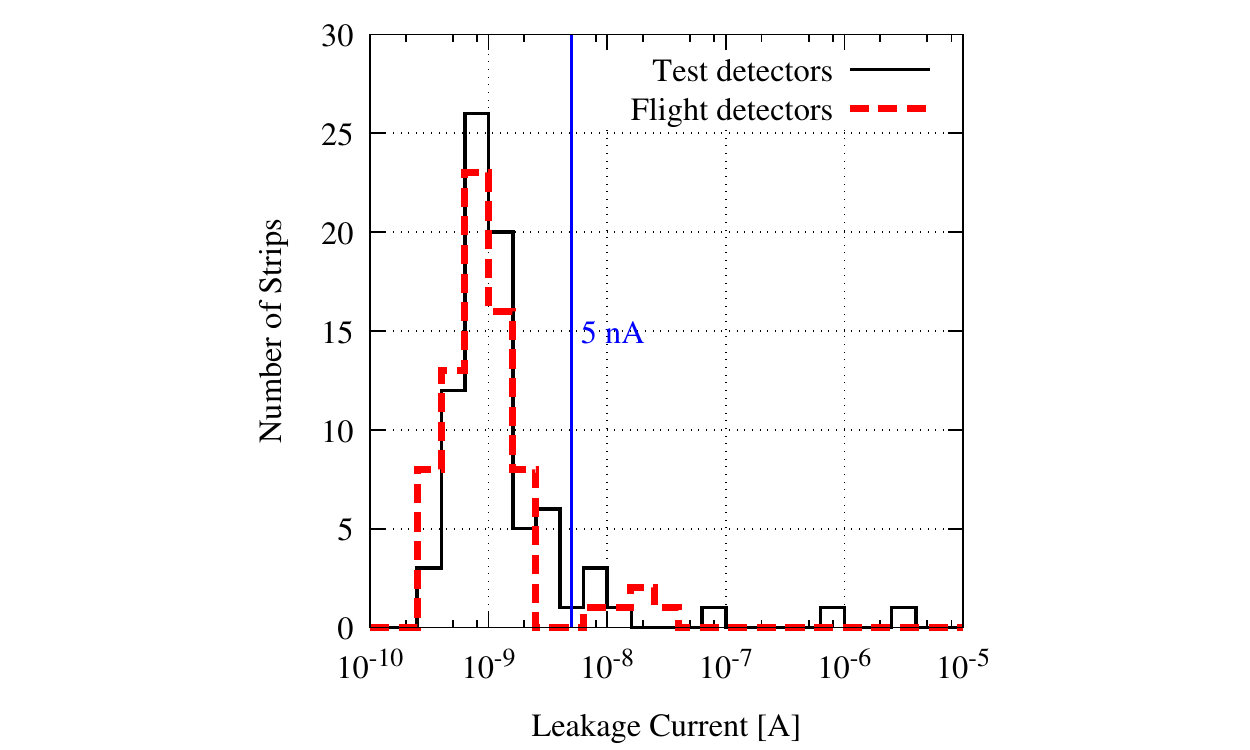}
		\caption{Histograms of the leakage current measured for
		all strips of the 10 test detectors (black line) and 10 flight detectors (red dashed line).
		The bias voltage and the temperature are set at 250~V and $-35^\circ$C, respectively.}
	\label{fig:hist}
	\end{center}
\end{figure}
Figure~\ref{fig:hist} displays histograms of the leakage current
measured for all strips of the 10 test detectors and 10 flight detectors.
The bias voltage and the temperature are set at 250~V and $-35^\circ$C, respectively.
In 80 strips of the 10 test (10 flight) detectors, 73 (75) strips have
the leakage currents below 5~nA.
Our acceptance criteria of the leakage current of each detector,
that at least 7 strips out of 8 must have leakage currents ${\leq}$5~nA,
was satisfied by 9/10 test detectors and 9/10 flight detectors.
All strips of the test and flight detectors pass our acceptance criteria
of the capacitance, ${\leq}$42~pF at 250~V at room temperature,
and we confirmed that the capacitances at $-35^\circ$C stays within 97-99\% of the room temperature value
for all 80 strips of the test detectors.
Capacitances of the flight detectors at cold temperature are not measured in our mass-production scheme
to cut the lead-time of the mass production.

\section{Conclusions}\label{sec:conclusion}
We have developed and established, for the first time,
a high-yield fabrication method for large-area Si(Li) detectors
operated at relatively high temperature.
Our 10~cm-diameter $p$-type crystal is verified
to have sufficiently low impurity concentrations and radial non-uniformities
to fabricate a uniform Li-drifted layer.
While a large-area Schottky barrier contact on the $p$-side can easily break down at $-35^\circ$C,
our optimized drift sequence retains a uniform undrifted layer on the $p$-side
with a thickness of ${\sim}100 \;\upmu\rm{m}$,
which drastically reduces bulk leakage current.
The guard-ring groove and optimized etching process are also confirmed
to effectively reduce the leakage current.\par
Our Si(Li) detector design has a sensitive layer with a ${\gtrsim}2.2$~mm depth,
${\gtrsim}90\%$ depth of the overall detector thickness of 2.5~mm,
and an overall diameter of 10~cm, with ${\sim}$9~cm of active area.
The detector is segmented into 8 readout strips and
the operating temperature is ${\sim}-40\pm5^\circ$C.
Based on the 20 mass-produced detectors we have produced,
the production yield is sufficiently high (${\sim}$90\%) at this point.
The detectors' leakage currents and capacitances are low enough
to achieve ${\lesssim}4$~keV energy resolution for 20$-$100~keV X-rays.
Good uniformity of the capacitance between strips indicates
that the detector volume is uniformly compensated by our Li drift.
This is also confirmed by copper staining on cross-sections of sample Si(Li) detectors.\par
The mass production of ${\gtrsim}1000$ GAPS Si(Li) detectors was started in late 2018
and will extend through early 2020.

\section*{Acknowledgements}
We thank SUMCO Corporation and Shimadzu Corporation for their co-operation in our detector development.\par
M. Kozai is supported by the JSPS KAKENHI under Grant No. JP17K14313.
H. Fuke is supported by the JSPS KAKENHI under Grants No. JP2670715 and JP17H01136.
K. Perez receives supports from the Heising-Simons Foundation,
the Alfred P. Sloan Foundation, and the ULVAC-Hayashi MISTI Seed Fund.
F. Rogers is supported through the National Science Foundation Graduate Research Fellowship under Grant No. 1122374.\par
This work was partially supported by the JAXA/ISAS Small Science Program in Japan and
the NASA APRA program through Grant NNX17AB44G in the US.\par
We are grateful to all GAPS collaborators.




\section*{References}
\bibliographystyle{elsarticle-num} 
\bibliography{bibfile}




\end{document}